\def\lt {{LaTe$_{3}$}}
\def\ef{{$E_F$}}
\def\a {{\AA$^{-1}$}}
\def\q {$q_{_{\rm{CDW}}}$}
 
\documentclass[preprint,onecolumn,amsmath,amssymb,aps,floatfix]{revtex4-2}
\usepackage{float}
\usepackage{wasysym}
\usepackage{graphicx}% Include figure files
\usepackage{dcolumn}% Align table columns on decimal point
\usepackage{bm}% bold math
\usepackage{multirow}
\usepackage{array}
\usepackage{booktabs}
\usepackage{amsmath}
\usepackage[normalem]{ulem}
\usepackage{physics}
\usepackage[dvipsnames]{xcolor}
\usepackage[section]{placeins}
\usepackage{comment}
\usepackage{xr-hyper}
\usepackage{hyperref}
\hypersetup{ colorlinks=true,citecolor=black, linkcolor=blue, filecolor=blue, urlcolor=black, pdftitle={Overleaf Example}}

\begin{document}
	\title {Charge density wave induced nodal lines in \lt}
\author{Shuvam Sarkar$^1$, Joydipto Bhattacharya$^{2,3}$,    Pampa Sadhukhan$^1$, Davide Curcio$^{4}$,  Rajeev Dutt$^{2,3}$,  Vipin Kumar Singh$^{1}$, Marco Bianchi$^{4}$,   Arnab Pariari$^{5}$, Shubhankar Roy$^6$,  Prabhat Mandal$^{5}$,  Tanmoy Das$^{7}$, Philip Hofmann$^{4}$,  Aparna Chakrabarti$^{2,3}$, Sudipta Roy Barman$^{1}$ }
\affiliation{$^1$UGC-DAE Consortium for Scientific Research, Khandwa Road, Indore 452001, Madhya Pradesh,  India}
\affiliation{$^2$Theory and Simulations Laboratory,  Raja Ramanna Centre for Advanced Technology, Indore 452013, Madhya Pradesh, India}
\affiliation{$^3$Homi Bhabha National Institute, Training School Complex, Anushakti Nagar, Mumbai  400094, Maharashtra, India}
\affiliation{$^4$Department of Physics and Astronomy, Interdisciplinary Nanoscience Center (iNANO), Aarhus University, 8000 Aarhus C, Denmark}
\affiliation{$^5$Saha Institute of Nuclear Physics, HBNI, 1/AF Bidhannagar, Kolkata 700 064, India} 
\affiliation{$^6$Vidyasagar Metropolitan College, 39, Sankar Ghosh Lane, Kolkata 700006, India}
\affiliation{$^7$Department of Physics, Indian Institute of Science, Bangalore,  560012, India} 

\begin{abstract}
	
	\lt\, is a noncentrosymmetric (NC)  material with time reversal (TR) symmetry in which the charge density wave  (CDW) is hosted by  the Te bilayers. Here, we show that \lt\, hosts a Kramers nodal line (KNL), a twofold degenerate nodal line that connects the TR invariant momenta in NC achiral systems, using angle resolved photoemission spectroscopy (ARPES), density functional theory (DFT), effective band structure (EBS) calculated by band unfolding, and symmetry arguments. DFT incorporating spin-orbit coupling (SOC) reveals that the KNL -- protected by the TR and lattice symmetries -- imposes gapless crossings between the bilayer-split CDW-induced shadow bands and the main bands. In excellent agreement with the EBS,  ARPES data corroborate the presence of the KNL and show that the crossings traverse the Fermi level. Furthermore,  spinless nodal lines -- entirely gapped out by the SOC -- are formed  by the linear crossings of the shadow and main bands with a high Fermi velocity. 
	
\end{abstract}
\maketitle
\section{Introduction}
Recent years have witnessed a rapid development in the understanding  of  the physics 
of   cooperative charge density wave (CDW) electronic state~\cite{Wang2022,%Raman on LaTe3 showing axil Higgs boson Nature, %Lin2021,
	Jiang2021,% Kagome system CDW NMat
	Shi2021,% CDW semimetal gaps out Weyl points making axion insulator NPhys
	Rettig2016,%Time dependent arpes on HoTe3 Ncomm
	Luo2022,%Kagome CDW NComm
	Yu2021,%Kagome CDW NComm
	Zong2019, %photo induced CDW in LaTe3 and TD ARPES NPhys
	Li2021,%CDW without acoustic phon anomaly PRX
songPRL2022,%CDW in monolayer heterostructure TiSe2 PRL
LvPRL2022,%Hysteretic transition in a CDW PRL
PanSDSharma2022,%topological CDW PRL %Pariari2021,Lei2020,brouet2004,
DSouza2012}. 
~In particular,  interplay of  the CDW electronic state  with the non-trivial topological phases provides 
an interesting platform for discovery of novel  quasiparticles  such as, axionic insulator~\cite{Shi2021,Gooth2019}, quantum spin-hall insulator~\cite{Qian2014},  fractional Chern insulator states~\cite{Polshyn2022}%Wilhelm2021} 
~and  manipulation of topologically protected states~\cite{Mitsuishi2020,Lei2021}. CDW  can drive topological phase transitions by modifying the  symmetry of the lattice, such as breaking the inversion symmetry~\cite{Hsu2021}. Interesting topological phases are frequently found in noncentrosymmetric materials, such as, nodal chain fermions~\cite{Bzdusek2016}, Dirac and Weyl fermions~\cite{Xia2020,Gao2018,Oh2019}, hourglass fermions protected by glide reflection~\cite{Leonhardt2021}, Kramers Weyl semimetal (KWS)~\cite{Chang2018} and  recently predicted Kramers nodal line (KNL) metal~\cite{Xie2021}. KNLs differ from the  Weyl nodal lines because they join two time reversal invariant momenta (TRIM) points and should appear in all achiral noncentrosymmetric time reversal symmetry (TRS) preserving systems~\cite{Xie2021}. For the subclass of nonsymmorphic symmetry, KNLs emerge from the $\Gamma$ points. Unlike the previously known nodal lines manifested by band inversion\cite{Yu2015}, the KNLs are robust under spin-orbit coupling (SOC) unless the protecting lattice symmetries such as TRS,  mirror, or roto-inversion symmetries are broken.    KNL fermions have been predicted to exhibit novel physical properties such as quantized optical conductivity~\cite{Xie2021}.  However, in this emerging field, to the best of our knowledge, the experimental evidence of KNL  is limited to the  work by Shang \textit{et al.}~\cite{Shang2022} who reported that  transition metal ruthenium silicides belong to this class  and exhibit unconventional  superconductivity based on muon spin spectroscopy and density functional theory (DFT). 

In recent years, multiple fascinating findings  in \lt~\cite{Wang2022,Zong2019,Kogar2020,Rettig2016,Zong2021,Pariari2021}
-- a member of the RTe$_3$ (R represents a rare earth element) family with highest CDW transition temperature of 670~K~\cite{Hu2014,Yumigeta2021} --  have  rekindled the scientific  interest of the community in this noncentrosymmetric TRS preserving  material. The detection  of  an axial Higgs boson mode in  LaTe$_3$ from Raman spectroscopy   has been related to unconventional  CDW excitation~\cite{Wang2022}. Photoinduced CDW state with topological defects in this material  has been discovered from ultrafast electron diffraction and related studies~\cite{Zong2019,Kogar2020}.    Transport studies have revealed that \lt\, possesses an unusually high non-saturated longitudinal magnetoresistance~\cite{Pariari2021}, which is similar to that of nodal line material~\cite{Singha2017}. In addition, \lt\, possesses very high carrier mobility~\cite{Pariari2021}, which, in conjunction with its high transition temperature, makes it a promising contender for next-generation electronics.   Theoretical calculation of the electronic susceptibility as well as experiments have shown that  $q$ dependent electron phonon coupling plays an important role in stabilizing the CDW state in RTe$_3$~\cite{Johannes2008,Eiter2013}.  While there are no DFT investigations, angle resolved photoemission spectroscopy (ARPES) measurements on \lt\, are also scarce in literature~\cite{Brouet2008,Zong2019}.  Using time resolved ARPES, Zong \textit{et al.}~\cite{Zong2019} examined light-induced melting of the CDW state of \lt. A previous research by Brouet \textit{et al.}~\cite{Brouet2008} demonstrated that the CDW-induced shadow bands hybridize with the main bands existing in the non-CDW state creating a CDW gap along a particular direction of the Brillouin zone (BZ). 

 In light of the fact that \lt\, is a noncentrosymmetric achiral material with TRS intact,  topological phases brought on by inversion symmetry breaking in conjunction with other symmetries may be anticipated in the CDW state. Here, from an in-depth study of the band structure of   \lt\, in the CDW state  by combining ARPES and  $ab-initio$ DFT using a realistic experiment-based structure, we establish  the existence  of a KNL in a CDW material for the first time.  It  originates from the interaction of the shadow band and the main band, and is hosted by the TRS and the lattice symmetries. Furthermore, spinless nodal lines that are entirely gapped out by  SOC are also identified.

\section{Results}

\subsection{Modulated structure of \lt\, in the CDW state}
\label{structure} 
\begin{figure}[tbh] %[H]
	\includegraphics[width=0.75\textwidth,keepaspectratio,trim={0cm 0cm 1cm 2cm},clip]{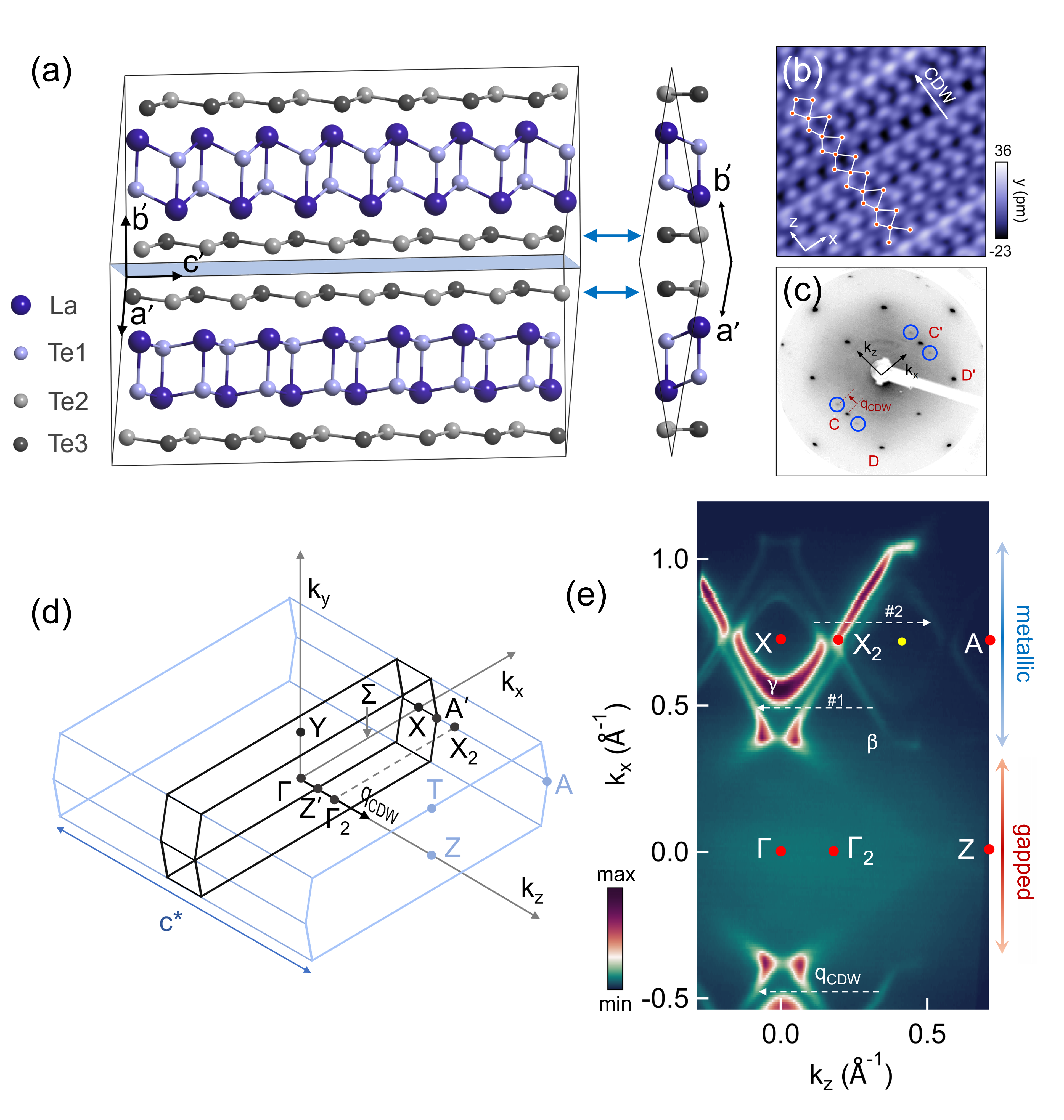}
	\caption{\textbf{Structure, Brillouin zone and the Fermi surface of \lt\, in the CDW state.}~(a)  The 7-fold (1$\times$1$\times$7) modulated primitive unit cell   of  LaTe$_{3}$ comprising of 56 atoms with  \q= $\frac{2}{7}$$c^*$  viewed perpendicular  (left) and parallel (right) to  the \textbf{c$^{\prime}$} direction. The lattice constants  are $a^{\prime}$= $b^{\prime}$= 13.256 \AA, $c^{\prime}$= 30.778~\AA~ with $\alpha^{\prime}$= $\beta^{\prime}$= 90$^{\circ}$, and $\gamma^{\prime}$=160.99$^{\circ}$.  The cleavage plane (light blue) occurs  in between the two weakly interacting Te layers. (b) High resolution scanning tunneling microscopy  topography image 	obtained with bias voltage of  0.2 V and a tunneling current of  0.4 nA. (c) Low energy electron diffraction   pattern  measured with 57 eV primary beam energy in inverted gray scale.  	(d)  The  CDW Brillouin zone   (BZ)   (black)  is shown within that of the non-CDW state (light blue). The   high symmetry points~\cite{Setyawan2010} are indicated in these respective colors. $\Gamma$, $Y$ and $X$ points coincide for both.     $\Gamma_2 X_2$ (dashed line)  is in the 2$ ^{\rm nd} $ BZ along $k_z$. (e) 	 The  Fermi surface (FS)  in the CDW state  measured by ARPES, where % using 24.4 eV photon energy.  
		~the length of the white dashed arrows (\#1, \#2)  that join  the shadow branches with the main branches of the FS represents \q.} 
	\label{cdw}
\end{figure}

Fig.~\ref{cdw}a shows a  7-fold modulated (1$\times$1$\times$7) supercell  structure of \lt.  It is derived from the experimental structure obtained from  single crystal x-ray diffraction (XRD)  at 100~K~\cite{Malliakas2006} (see \textit{Methods}).  Its space group  is $C2cm$ ($SG$~\#40), which is same as that reported for the average structure~\cite{Malliakas2006}.  \lt\, is made up  of two main structural units:   La-Te1 corrugated slabs and  Te2-Te3   bilayer. The latter,  highlighted by blue double-sided arrows, are weakly coupled by van der Waals interaction.  The Te bilayer of the 7-fold modulated structure hosts the CDW with  \q\,= $\frac{2}{7}$$c^*$= 0.286$c^*$, where c$^*$ is the reciprocal lattice vector along $k_{z}$ in the non-CDW state. This value is close  to the experimental values determined  from ARPES (subsection \ref{shadowband}) and reported in previous literature~\cite{Malliakas2006,DiMasi1995}.  Moreover, the amplitude of the CDW modulation  is nearly similar to that from  XRD~\cite{Malliakas2006}, see  Supplementary Figs.~S1a,b.%\ref{supp-qCDW_fit}a,b. 
	
	The CDW modulation has been directly observed in the \lt\, crystal studied here from a  high resolution scanning tunneling microscopy (STM) topography image.  In Fig.~\ref{cdw}b, the modulated  Te net formed by connecting  the Te atoms (orange circles) is shown by white lines.   Low energy electron diffraction (LEED) pattern also  shows  CDW related satellite spots (encircled) besides the (1$\times$1) spots  (Fig.~\ref{cdw}c). Satellite spots are also  observed in the  Fourier transform  of the  STM image, see Supplementary Note~1 and Fig.~S2.%\ref{supp-STM_supple}. 
	
The experiment based  7 fold  modulated structure discussed above has been used for the DFT calculations.  The  corresponding BZ  is shown in Fig.~\ref{cdw}d inscribed within the non-CDW BZ.  Since the BZ is related to the ordering of the lattice constants of the conventional cell, 	its comparison with the primitive unit cell of the 7-fold structure    is shown in Supplementary Figs.~S3a,b.%\ref{supp-nonCDW_cell}a,b. 
~In our notation,  the horizontal plane in the BZ is represented by $k_x$-$k_z$  and \q\, is oriented along $k_z$.  The CDW BZ, containing features such as the primitive reciprocal lattice vectors and all the pertinent high symmetry points and directions as well as their coordinates, is depicted in Supplementary Fig.~S3c.%\ref{supp-nonCDW_cell}c. 

\subsection{Crossing of the bilayer-split shadow and main bands from ARPES}
\label{shadowband} 
\begin{figure*}[tbh]
	\includegraphics[width=1.05\textwidth,keepaspectratio,trim={2cm 0cm 2cm 2cm },clip]{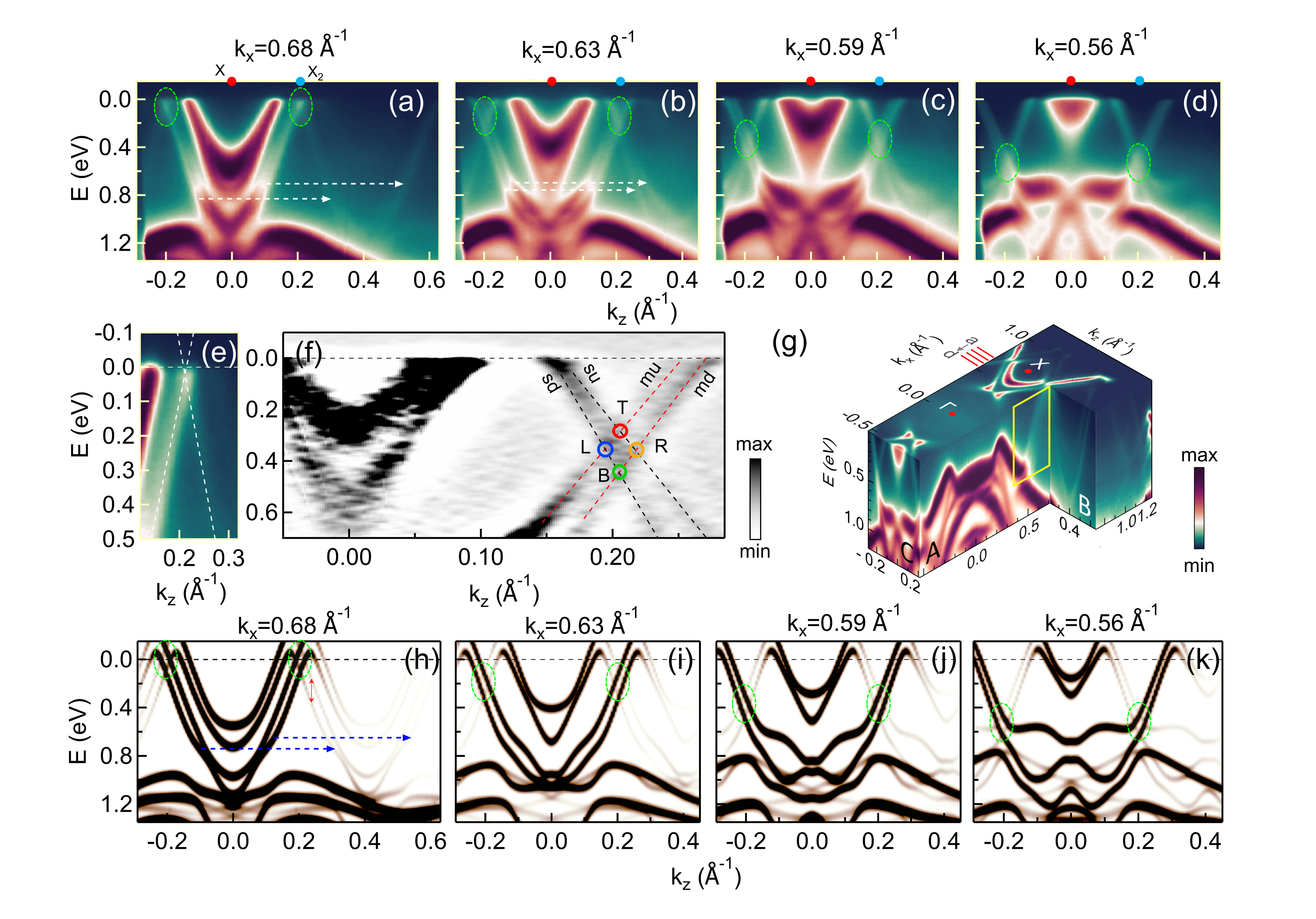}
	\caption{\textbf{Crossings between the bilayer-split  shadow  and the main bands.}~ $E(k_z)$ ARPES intensity plots measured at 100~K %in the range -0.25$\leq$$k_z$$\leq$0.63\,\,\a, 
		~for $k_x$=  (a) 0.68, (b) 0.63, (c) 0.59, and (d) 0.56 \a. The  crossing regions are highlighted by  green dashed ovals.  (e)  A zoomed view of the near \ef\, region of  panel \textbf{a} around $k_z$= 0.21\,\a, the white dashed lines are obtained by curve fitting (see \textit{Methods}).  (f)  Curvature plot of a part of panel \textbf{c} shown in an expanded scale:   the crossings indicated by  $L$, $B$, $R$ and $T$ are shown by blue, green, yellow and red  circles, respectively.  The $T$ and $B$ crossings occur at the $X_2$ point. %, i.e., the $X$ point of the 3rd BZ (panel \textbf{d} of Fig.~\ref{cdw}). 
		~  The dashed red (black) line representing the main (shadow) band  is obtained by curve fitting as in panel \textbf{e}.   (g)  Cut \textit{A} at $k_z$= 0.216 of an $E$-$k_z$-$k_x$  ARPES intensity plot shows the  dispersion of the   $R$ crossing and the $mu$ and $sd$ bands  within the yellow rectangle. %The corresponding  curvature plot is shown in Fig.~\ref{crossing}f. 
		A shadow band is observed in the cut \textit{B}.  (h-k) The effective band structure (EBS)  along $k_z$ obtained by band unfolding at similar $k_x$ values  as in panels \textbf{a-d}, respectively. The band crossing regions  are highlighted by green dashed ovals.  }
	\label{effcross} 	
\end{figure*}

An	$E(k_z)$ ARPES  intensity plot in  a generic direction parallel to $\Gamma Z$  at $k_x$= 0.68~\a\,  i.e., near the $X$ point (the length of $\Gamma X$ being 0.737~\a) is shown in Fig.~\ref{effcross}a (see \textit{Methods} for the experimental details). 
 The direction of the measurement is  shown by a red line denoted by ``a" in  Fig.~\ref{effcross}g  . It shows  two main bands (inner and outer) centered around $\Gamma$ that cross \ef\, at $k_z$= $\pm$0.15 and $\pm$0.21~\a, respectively. The outer main band disperses down to  binding energy ($E$) of about 1.25 eV, while the inner band has a nearly flat bottom at \AC0.8 eV. It is interesting to note that a relatively weaker band centered around $k_z$= $\pm$0.41\,\a\,  is a  replica of the main band shifted by  \q (= 0.28$c^*$= 0.41\,\a),  as shown by two horizontal white dashed arrows.  This replica band crosses \ef\, at $k_z$= 0.21 and 0.61\,\a\, and  disperses down to $E$\AC1.25~eV.  No shift in its position  along $E$ compared to the main band is observed, which indicates that it is  related to the initial state CDW superlattice~\cite{Komoda2004}.  It has been referred to in literature as the shadow band~\cite{Brouet2008,Komoda2004,Mans2006}.  The   signature of the shadow band is also evident in the Fermi surface (FS) shown in Fig.~\ref{cdw}e, where shadow FS branches appear at a separation of \q\, from  the main branches, as shown by white dashed arrows in the metallic region around the $X$ point parallel to the $\Gamma Z$ direction.  In the Supplementary Note 2, a discussion about the FS and  \q\, obtained from the average separation of the shadow and main branches are provided. \q= 0.28$\pm$005$c^*$ determined in this way is close to  $\frac{2}{7}$$c^*$ and is in excellent agreement with the values obtained from  STM and LEED, see Supplementary Note~1.

In Fig.~\ref{effcross}a, The shadow  and the main bands resemble an inverted ``V"   and meet each  other  close to \ef\,  at $k_z$= $\pm$0.21\,\a (highlighted by a green dashed oval). This region is shown in an expanded scale in Fig.~\ref{effcross}e, where  the white dashed lines suggest a possible crossing of  nearly  linear bands at \ef. The Fermi velocities of these  bands,  determined using the expression  ($\frac{1}{\hslash}$$\frac{dE}{d\vec{k_z}}$),  turn out to be 1.2$\pm$0.05$\times10^{6}$ and 1$\pm$0.05 $\times10^{6}$ m/s  for the shadow and main bands, respectively.  These values are comparable to graphene (1$\times10^{6}$ m/s~\cite{Novoselov2005}), indicating  large mobility of \lt\,  in agreement with recent report from  Hall conductivity measurements~\cite{Pariari2021}. 
   
    To further investigate these bands and their possible crossing,   ARPES was performed  over a  range of $k_x$ out of which three representative plots (along the red lines parallel to ``a" up to ``d" in Fig.~\ref{effcross}g)  are shown in Figs.~\ref{effcross}b-d. Interestingly, as $k_x$ decreases, both the main and the shadow bands spread out in $k_z$ and  the  crossings -- highlighted by  green dashed ovals --  shift below \ef\, to larger $E$. However, note that their $k_z$ position remains  unchanged  around $X_2$ $\approx$ $\pm$$\frac{1}{2}$\q.  In Figs.~\ref{effcross}b-d,  splittings in both the shadow and the mains bands are observed, which  are   related to the coupling  between the two adjacent  weakly coupled Te  layers. This is the   bilayer splitting that  has been reported  in   other RTe$_3$ members% such as SmTe$_3$, CeTe$_3$ and YTe$_3$
   ~\cite{Gweon1998,brouet2004,Brouet2008}, bilayer graphene~\cite{Ohta2006},  and  cuprate superconductors~\cite{Feng2001}. %(Bi2Sr2CaCu 2O8+δ)
 ~  Here, we find that the splitting increases with $E$, e.g., it varies from 0.016 to 0.03~\a\, as $E$ increases from 0.1 to 0.5 eV  (Fig.~\ref{effcross}c).    Evidence of varying bilayer splitting is also observed in the FS (see  Supplementary Note~2). 

Both the bilayer-split shadow bands - denoted by \textit{su} (\textit{sd}) for smaller (larger) $E$, see Fig.~\ref{effcross}f - are shifted by \q\, from the corresponding main bands (denoted by \textit{mu} and \textit{md}). This is shown by the white dashed  horizontal arrows in Figs.~\ref{effcross}a,b. This  leads to formation of  four  crossings between the  main and the shadow bands, as  shown  in Fig.~\ref{effcross}f.  These are also visible in a constant energy   $k_z$-$k_x$  isosurface through the band crossing  \textit{e.g.} at $E$=  0.5 eV  and a stack of momentum distribution curves (MDCs), see Supplementary Figs.~S8a,b.%\ref{supp-quartet_isosurface}a,b. 

The crossings are denoted by  \textit{left} (\textit{L}, blue circle, crossing of \textit{mu} and \textit{sd}  i.e., $mu \otimes sd$),  \textit{right} (\textit{R}, orange circle, $md \otimes su$),  \textit{top} ($T$, red circle, $mu \otimes$$su$), and \textit{bottom} (\textit{B}, green  circle, $md \otimes sd$) (Fig.~\ref{effcross}f). %srbfinal7: It is important to note  that  the bands are linear around the crossing points, and the velocities calculated from their slopes  are similar to that obtained at \ef\, in Fig.~\ref{effcross}a. 
~ While Figs.~\ref{effcross}a-d show the band dispersion at discrete $k_x$ values, we show a continuous dispersion of  the crossings, for example the $R$   in cut \textit{A} of Fig.~\ref{effcross}g.  Here,  the $E(k_x)$  at $k_z$=0.216\,\a\,  shows the loci of the $R$ crossing within the yellow rectangle. 
  
  ARPES with different photon energies shows negligible  $k_y$ dependence  of the crossings. For example at   k$_x$= 0.58~\AA$^{-1}$,  Supplementary Figs.~S9a-d%\ref{supp-ky_crossing}a-d  
  ~show that their position (highlighted by green dashed ovals)  remains almost unchanged with  $k_y$.   This  is summarized in  Fig.~S9e%\ref{supp-ky_crossing}e 
  ~ through a  $k_y$-$k_z$ map at $E$= 0.39 eV,  where the two crossings at $k_z$= $\pm$0.2~\AA$^{-1}$ (marked by yellow arrows) show almost no change with $k_y$, indicating the quasi-2D nature  of \lt.

\subsection{Effective band structure from DFT compared to ARPES}
\label{effbandstr} 

Multiple  crossings indicated by ARPES is a surprising result since   hybridization gap is generally expected  at the band touching points between the  Bloch states connected by \q~\cite{Yang2010,brouet2004}. To understand this, we have performed  DFT calculations using the  modulated  structure discussed in subsection \ref{structure}. 

 The $E(k_z)$  bands  calculated  at different $k_x$ as in the experiment  are shown in an extended zone scheme spread over the non-CDW BZ  in Supplementary Fig.~S10.%\ref{supp-Raw_band_XS_para}. 
  ~Many bands are observed in the CDW state  due to band folding. The complexity of these bands hides the influence of the CDW on the electronic bands and impedes their interpretation and direct comparison with ARPES. So, an effective band structure (EBS)   has been calculated by unfolding the bands in the  non-CDW BZ~\cite{Ku2010,Popescu2010,Herath2020}.  In  Figs.~\ref{effcross}h-k,  EBS shows the distribution of  states as a function of their energy and momenta with appropriate spectral weight, where broadening similar to the experiment has been applied.  Variations in the spectral weight  is evident resulting in  dissimilar EBS in the different CDW BZs.  This complexity in a CDW system is discussed  further in subsection \ref{knlarpes}.  

The importance of the EBS calculation is that despite the CDW effect being small with the modulation  amplitude  only  \AC4.5\%  	of the average Te2-Te3  distances, the  EBS    reveals  the  shadow bands 	that are  shifted from the  main bands by \q\, (blue dashed arrows) in  excellent agreement with ARPES (compare Figs.~\ref{effcross}h-k with Figs.~\ref{effcross}a-d).  	In contrast, the bands in the non-CDW state,  where obviously the CDW amplitude is zero, shows only the main bands (Supplementary Fig.~S11a).%\ref{supp-XS_nCDW}a).  
	~An analysis of the orbital character establishes that the shadow bands (as well as the main bands)  are of predominantly  in-plane Te $p_x$-$p_z$  character, see the Supplementary Figs.~S12a-f.%\ref{supp-orbital_projection}a-f. 
~The out-of-plane  $p_y$ character becomes slightly significant only for $E>$0.6~eV, and the contributions from the  La-Te1 block  is negligible.    However, transfer of electrons from La to the Te net  determines  the band filling in the Te2-Te3  layer and this has been calculated  using the Bader charge analysis (see Supplementary Note 3 including Refs.~\onlinecite{bader1985,Bhattacharya2023, SarkarAIP2020}).  

From Fig.~\ref{effcross}h, we find that the outer branch of the shadow band  disperses %linearly 
~towards \ef\, and crosses the main band  around $k_z$= $\pm$0.21 \a\, (highlighted by a dashed green oval). A  bunch of bands that   disperse weakly  between 1.15 to 1.6 eV are  observed in both theory and experiment at similar $E$.  A bilayer splitting of  0.17-0.2 eV (red double arrow in Fig.~\ref{effcross}h) is observed in the EBS  with $L$ and $R$ crossings close to \ef, while $T$ ($B$) appears above (below) \ef. However, these are not visible separately in Fig.~\ref{effcross}a, indicating that possibly  at this $k_x$ the experimental bilayer splitting is smaller compared to the DFT  value. The  inner branch of the shadow band that disperses to about 0.6 eV is more prominent in the EBS, but  also has its counterpart in the curvature plot of ARPES  (see Supplementary Fig.~S11b). In Figs.~\ref{effcross}i-k, for progressively smaller $k_x$ values, EBS portray  all the four crossings at similar $E$-$k_z$ observed in ARPES (compare Figs.~\ref{effcross}i-k with Figs.~\ref{effcross}b-d). The agreement  is also good at larger $E$:  the main bands around 0.8 eV become flatter and move to lower $E$. 

Important to note is that from ARPES  the crossings appear to be gapless (Fig.~\ref{effcross}a-f) within the experimental and lifetime broadening. Although this is supported by the EBS (Fig.~\ref{effcross}h-k),  it should be noted that the calculations have been performed with a $k$ step size ($\delta$$k_z$) of  6$\times$10$^{-3}$\a, %5.95x10^-3 A-1
~and with broadening comparable  to ARPES, which might  conceal the potential presence of minigaps.  In the subsequent subsections \ref{wosoc} and \ref{soc}, we show  the DFT bands calculated with smaller $\delta$$k_z$ to probe these crossings further.

\subsection{Spinless nodal lines formed by the $L$ and $R$ crossings}
\label{wosoc}

\begin{figure*}[tbh]
	\includegraphics[width=\textwidth,keepaspectratio,trim={3.5cm 0cm 1cm 0cm },clip]{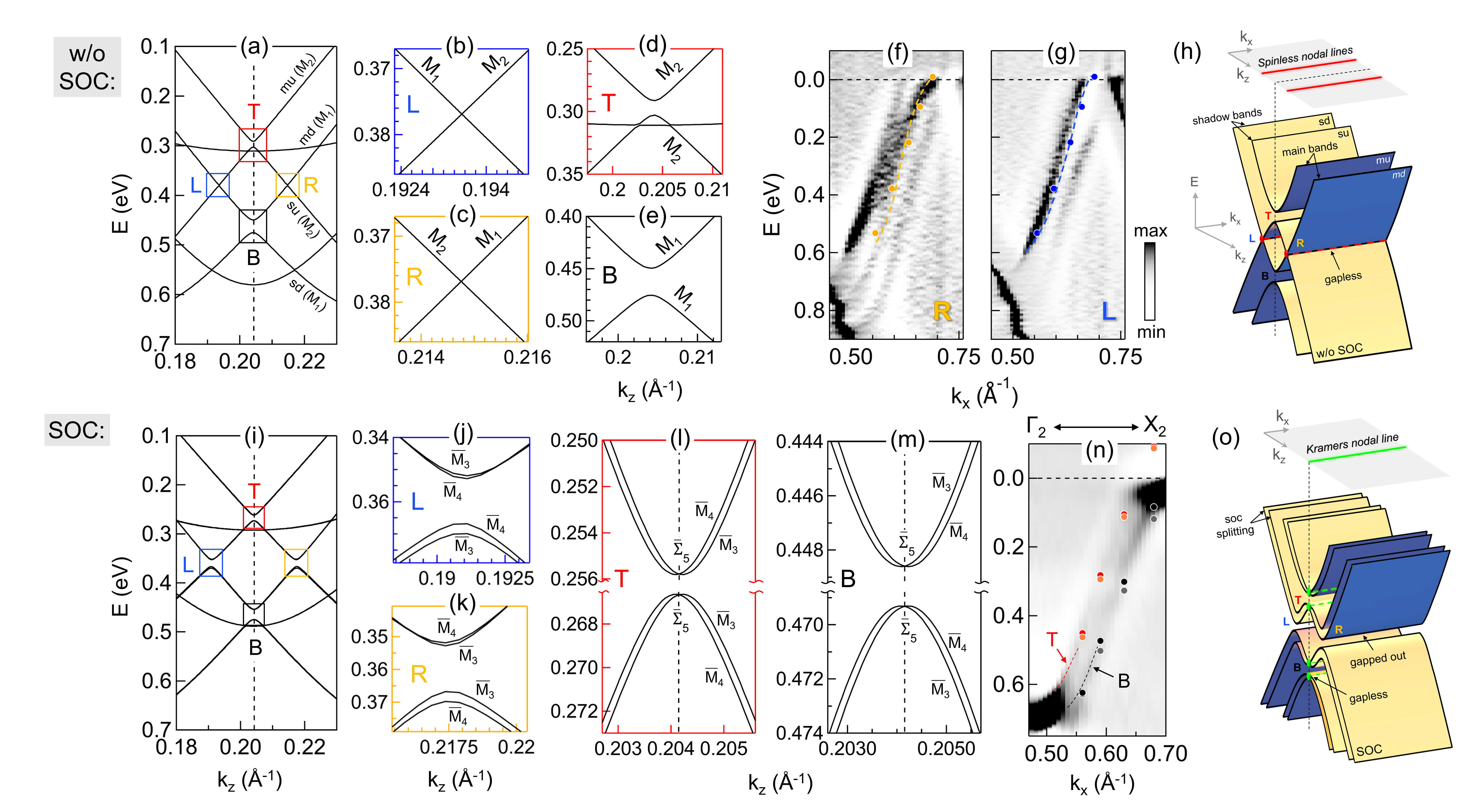}
\caption{\textbf{Band crossings and the Kramers nodal line.}~(a) $E(k_z)$ band dispersion from DFT 	at $k_x$= 0.59\,\a\, without (w/o) SOC. The irreducible representations (irreps) are shown.  The vertical dashed line represents the $k_z$ point on  $\Gamma_2 X_2$ i.e., the $\Sigma$ line.  Zoomed  colored rectangles of panel \textbf{a} show the   bands  around (b) $L$ ($mu \otimes sd$),  (c) $R$ ($md \otimes su$), (d) $T$ ($mu \otimes$$su$), and (e) $B$ ($md \otimes sd$). Comparison of the $E(k_x)$ ARPES intensity plot with DFT for the following crossings: (f) $R$ at $k_z$= 0.229~\a and (g) $L$ at $k_z$= 0.196~\a. The positions of these crossings obtained from DFT (orange and blue filled circles for $R$ and $L$, respectively) are superimposed.  
(h) A schematic representation of the  gapless $L$ and $R$ crossings in the $E$-$k_z$-$k_x$ space (red dashed lines) and their projection on the $k_x$-$k_z$ plane showing  the spinless nodal lines  (thick red lines on both sides of the $\Sigma$ line).  
(i-m) Same as panels \textbf{a-e} except that the calculations are performed with SOC.  (n)  $E(k_x)$ ARPES intensity plot at $k_z$= 0.204\,\a\, (dashed curves here and in panels \textbf{f} and \textbf{g} serve as guide to the eye) compared with the positions of the crossings enforced by the KNL obtained from DFT for  $T$ (red, light red circles) and $B$ (black, gray circles).  (o) A schematic representation of  the four  crossings (green dashed lines) related to the upper and lower branches of $T$ and $B$. The Kramers nodal line  appears along the $\Sigma$ line and is denoted by a green thick line on the $k_x$-$k_z$ plane. The gapped out spinless nodal lines at $L$ and $R$ are also shown. The energy dispersion along $k_x$ is not shown here as well as in panel \textbf{h}.}
\label{crossing}
\end{figure*}

The $E(k_z)$  DFT bands in the crossing region with small $\delta$$k_z$ (= 5$\times$10$^{-4}$\a)  in absence of SOC indeed  indicate that the $L$  and  $R$ crossings are   gapless (Fig.~\ref{crossing}a). % at  $k_x$= 0.59 \a.  
  This  is reconfirmed in Figs.~\ref{crossing}b,c by  the bands calculated with even smaller  $\delta$$k_z$ (=  1$\times$10$^{-5}$\a). It is interesting to note that these crossings occur at  generic points of the BZ in the $k_x$-$k_z$ plane and  the bands involved are linear (Figs.~\ref{crossing}a-c).  ARPES intensity plots in Figs.~\ref{effcross}c,f   at this $k_x$   also demonstrate the linearity of the  bands around $L$ and $R$ in excellent agreement with DFT. The velocities calculated from their slopes  are similar to that obtained at \ef\, from Fig.~\ref{effcross}e.  Although these bands originate from the in-plane $p$ orbitals with small difference in  contributions from  the $p_x$ and $p_z$ orbitals as shown in Supplementary Figs.~S12a-f,%\ref{supp-orbital_projection}a-f, 
  ~  \textit{md} and \textit{sd} bands belong to  $M_1$ irreducible band representation (irrep), while \textit{mu} and \textit{su} belong to   $M_2$ irrep. Thus both $L$ and $R$  crossings are formed by bands  belonging to  different irreps (Figs.~\ref{crossing}a-c).

 The crossings disperse  with $k_x$ and  at larger $k_x$= 0.685\,\a\, compared to 0.59\,\a,   $L$ and $R$ crossings traverse  the \ef\, (Supplementary Fig.~S13a-c).%\ref{supp-ef_crossing}a-c). 
~In fact, calculations for a series of $k_x$ values  establish that the gapless linear crossings occur over an extended range of the $E$-$k$ space  (Supplementary Fig.~S14).%\ref{supp-dense_crossing}).  
~ The loci of each crossing  form  a  continuous curve in the momentum space that has been referred to as a nodal line.   A direct comparison of the ($E, k_x$) cuts from ARPES (Figs.~\ref{crossing}f,g for $R$ and $L$, respectively) show that the positions  of both the  nodal lines are in excellent agreement with DFT. Both disperse  between ($E$, $k_x$) = (0 eV, $\sim$0.7\,\a) to (\AC0.6 eV, $\sim$0.5\,\a) with $k_z$ at 0.195  and 0.22\,\a\, for $L$ and $R$, respectively. Thus, the crossings appear within an energy window of $E$\AC0.6 eV to the \ef. Their projections in the $k_x$-$k_z$ plane form a  pair of approximately parallel  nodal lines that are 0.2\,\a\, in length and appear at a separation of $k_z$\AC0.02\,\a\,    in a general direction on this  plane  (red solid lines in Fig.~\ref{crossing}h). These appear parallel to but on either sides of  $\Gamma_2 X_2$ that occurs at  k$ _z $= 0.204\,\a. These nodal lines  are formed by two fold crossings of nondegenerate bands in the absence of spin and so are referred to spinless nodal lines. 
 
With inclusion of  SOC, it is intriguing to find that  both the $L$ and $R$ crossings (and hence the nodal lines) are entirely gapped out by minigap of 14-17 meV into upper and lower branches, as shown in Figs.~\ref{crossing}i-k  and Supplementary Figs.~S13g,h.%\ref{supp-ef_crossing}g,h. 
~A schematic is shown in Fig.~\ref{crossing}o. The gap is formed by hybridization of bands belonging to same one dimensional double group irrep  involving both $\overline{M_3}$ and $\overline{M_4}$. 
 
 Note that noncentrosymmetry should lift the SU(2) spin degeneracy.  This is visible through the spin splitting of both the upper and lower branches in Figs.~\ref{crossing}j,k.  These spin-split bands belong to different irreps $\overline{M_3}$ and $\overline{M_4}$ and do not cross each other. The splittings for both $L$ and $R$ are  $k$-dependent, it is  maximum  at the extrema (\AC3 meV for lower and 1 meV for the upper branch in  Figs.~\ref{crossing}j,k) and decreases away from it.

\subsection {Evidence of Kramers nodal line from the $T$ and $B$ crossings}
\label{soc}

In the case of $T$ and $B$, the shadow and main bands belong to the same irrep, leading to hybridization-related minigaps  without SOC and the creation of upper and lower branches  (Figs.~\ref{crossing}a,d,e). The inclusion of SOC provides a fascinating outcome:  the spin-split bands in  both the upper  and lower branches of  $T$ and $B$   exhibit  gapless crossings (Figs.~\ref{crossing}l,m), which contrasts the bands at $L$ and $R$ (Figs.~\ref{crossing}j,k).   Each branch of $T$ and $B$ crosses at the  same $k_z$= 0.204~\a\, (dashed lines in Figs.~\ref{crossing}i,l,m) resulting in four crossings.  This value of $k_z$ is  special, since it falls on the $\Gamma_2 X_2$ line  of the CDW BZ  (i.e., $\Gamma X$ or  the $\Sigma$ line) (Fig.~\ref{cdw}d).  At $k_x$= 0.59\,\a, the four crossings appear at different $E$:  0.256 (0.267) eV for the upper (lower) branch of $T$, and 0.448 (0.469) eV for upper (lower) branches of $B$. 

These gapless crossings are  observed over a range of $k_x$, as in the case of $L$ and $R$ (subsection \ref{wosoc}). For example, at $k_x$= 0.685~\a, the crossings  disperse to smaller $E$ and in this case  both the branches of $T$ ($B$) are above (below) the \ef\, (Supplementary Fig.~S13f-h).%\ref{supp-ef_crossing}f-h). 
~Significantly, the crossings always appear at   same $k_z$  i.e., along the $\Sigma$ line, suggesting that the crossings may be enforced by the lattice symmetries along this direction.
 
 The band irreps  with SOC shown in Fig.~\ref{crossing}l,m are as follows: the crossing belongs to a double valued irrep that is  two dimensional ($\overline{\Sigma}_5$), while away from it, the  spin-split bands have one dimensional ($\overline{M}_3$ or $\overline{M}_4$) representation.  Significantly, the $\Sigma$ line that emerges from the $\Gamma$ point has the little group that is isomorphic to $C_{2v} $  point group and has symmetries such as the  two fold rotation about the $k_x$- axis denoted by $\lbrace$2$_{100}$$\vert$0,0,0$\rbrace$,  glide reflection  perpendicular to the $y$ axis in the $k_x$-$k_z$ plane  followed by a translation  of $\frac{1}{2}c$  given by $\lbrace$$m_{010}$$\vert$0,0,$\frac{1}{2}$$\rbrace$ and an off-centered mirror perpendicular to the $k_z$ axis $\lbrace$$m_{001}$$\vert$0,0,$\frac{1}{2}$$\rbrace$.  $\Gamma$ is a TRIM point, %($k$= −$k$ + $g_i$, $g_i$ is a reciprocal lattice vector), 
 ~where according to the Kramers theorem, each  band is at least doubly degenerate.  The little group of $\Sigma$ is related to that of $\Gamma$ by the compatibility relations.  We find that  $\Gamma$  and $\Sigma$ are represented by  two dimensional double-valued  irreps: $\overline{\Gamma}_5$ and $\overline{\Sigma}_5$, respectively. These representations are similar, as shown in Supplementary Table~S2%\ref{supp-irrep}
 ~\cite{Elcoro2017}.  The $\Sigma$ line  passes through the $X$ point with co-ordinates (0.257, 0.257,0) and meets  the TRIM point $Y_2$ (0.5, 0.5, 0) in the next BZ (see Supplementary Fig.~S15).%\ref{supp-band_connectivity}). 
 ~$Y_2$  also belongs to double-valued    two dimensional irrep  %$\overline{Y}_5$ 
 ~$\overline{Y_2}_5$ that is same as $\Gamma$ and $\Sigma$ (Supplementary Table~S2).%\ref{supp-irrep}). 
 ~The condition that at the TRIM points  the representations should be time reversal invariant is satisfied since both $\overline{\Gamma}_5$ and $\overline{Y_2}_5$ %$\overline{Y}_5$ 
 ~are  pseudo-real~\cite{Elcoro2017}.  Thus,  $\overline{\Gamma}_5$-$\overline{\Sigma}_5$-$\overline{Y_2}_5$%$\overline{Y}_5$
 ~is able to support two fold degeneracy of the bands along  $\Gamma X Y_2$ i.e., the $\Sigma$ line. This is a Kramers nodal line (KNL) that occurs along the $\Sigma$ line in the mirror-invariant $k$-plane in presence of TRS, and the additional rotational symmetry  constrains the KNL along a high symmetry direction~\cite{Xie2021}. Xie~\textit{et al.}~\cite{Xie2021} have proved the existence of KNL along the $C_2$ rotational axis ($\lbrace$2$_{100}$$\vert$0,0,0$\rbrace$), which in our case is along $k_x$ that lies in the $k_x$-$k_y$ mirror plane ($\lbrace$$m_{001}$$\vert$0,0,$\frac{1}{2}$$\rbrace$).  In Fig.~\ref{crossing}o, the  projection of the loci of the  crossings of the four pairs of bands  in the $E$-$k_z$-$k_x$ space on the $k_x$-$k_z$ plane   shows the KNL (thick green line) that enforces the crossings.  
 
The bands in a plane that cuts the KNL perpendicularly -- as is the case in Fig.~\ref{crossing}i -- have been described in the literature  as  two-dimensional massless Dirac Hamiltonian with the Berry curvature concentrated at the crossing~\cite{Xie2021}. These authors demonstrated that the Berry phase around a KNL is quantized as m$\pi$ mod 2$\pi$. In case of quadratic and cubic dispersion of these bands, the crossing has been dubbed as a higher-order Dirac point~\cite{Xie2021}.  In the present case, both the upper and lower branches of $T$ and $B$ exhibit quadratic dispersion very close to the crossings,  as is evident in Figs.~\ref{crossing}l,m.  Consequently, the gapless  crossings in \lt\, that are associated with  the KNL are  higher-order Dirac points.

 \subsection{ARPES and DFT along the KNL and other directions}
\label{knlarpes}

\begin{figure}[t]
	\includegraphics[width=0.8\textwidth,keepaspectratio,trim={1.1cm 0cm 0cm 0cm },clip]
	{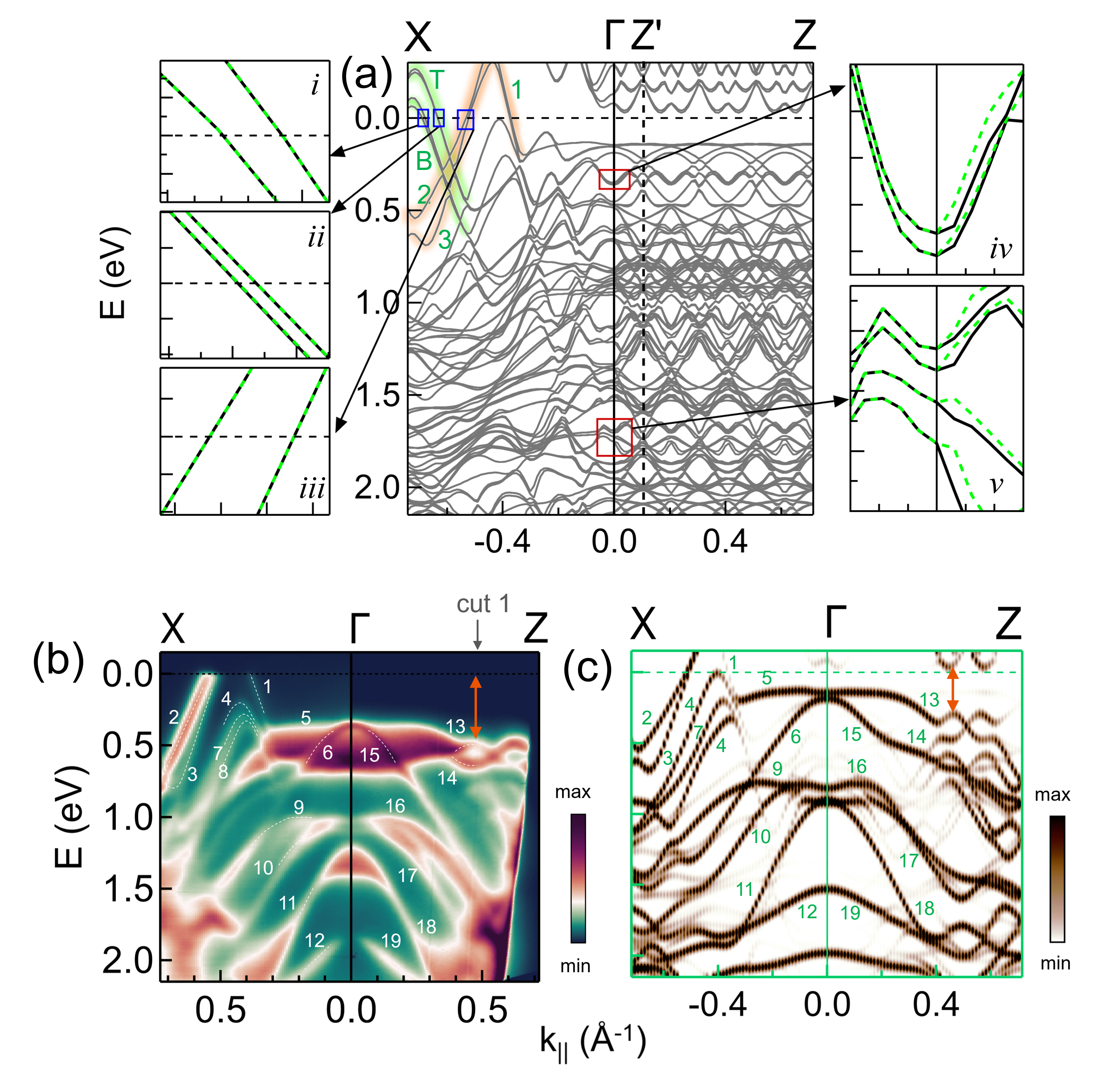}	%{Raw_XGZ_v1}
	\caption{\textbf{DFT and ARPES along and perpendicular to the KNL.}~(a) The band structure of \lt\, along   $X \Gamma Z^{\prime}Z$ in the CDW state. On the left,  zoomed regions from the blue rectangles show the two fold degeneracy (dashed-green and black) of the bands along the KNL.  On the right,  zoomed regions from  the red rectangles show the spin-splitting along the $\Gamma Z$ direction in contrast to $\Gamma X$. (b)   ARPES intensity plot   measured using photon energy  of 24.4 eV and (c)  the  EBS obtained from  DFT   towards  $\Gamma X$  and $\Gamma Z$.  } 
	\label{bandsxgz}	
\end{figure}

In Fig.~\ref{bandsxgz}a, the $E(k_x)$ bands calculated  along the KNL  (i.e., $\Gamma X$)  show  two pairs of  degenerate bands  related to the $T$ and $B$ crossings, highlighted by green shading in the  insets \textit{i, ii}.  The ARPES intensity plot in Fig.~\ref{crossing}n represents these bands that are also identified by the degenerate crossings of bands in the perpendicular $E(k_z)$ direction (Figs.~\ref{crossing}l,m) enforced by the KNL. The existence of the KNL between 0.5 to 0.7~\a\, along $k_x$ is affirmed by ARPES through an excellent agreement of the positions of the crossings obtained from DFT  (filled circles) that are overlaid on the experimental data in Fig.~\ref{crossing}n. The agreement is also evident in the Supplementary Fig.~S16,%\ref{supp-arpes_bands_superposed},  
~where Fig.~\ref{crossing}i is superposed on Fig.~\ref{effcross}f. However, the energy separations between the upper and lower branches of both $T$ and $B$ are too small to be resolved by ARPES. Similarly, the quadratic nature visible in a tiny $k_z$ range shown  in  Figs.~\ref{crossing}l,m is not distinguished, see the regions enclosed by the red and black rectangles in Supplementary Fig.~S16.%\ref{supp-arpes_bands_superposed}. 
~Nonetheless, ARPES is consistent with DFT,  and both show that   the crossings disperse from  $E$$\sim$0.6 eV at  $k_x$$\sim$0.55\,\a\, and traverse  the \ef\,  at $k_x$= 0.65-0.7\,\a.

We find that every band along $\Gamma X$ is degenerate due to the double degeneracy enforced by the KNL (see for example the  insets \textit{i-v} of Fig.~\ref{bandsxgz}a) and belong to two dimensional $\overline{\Sigma}_5$ irrep.  
Besides the $T$ and $B$ related bands discussed above, there are other  bands that cross \ef\, in Fig.~\ref{bandsxgz}a. These bands  -- numbered as \textit{1}, \textit{2} and \textit{3}  and highlighted by orange shading -- have not been observed in the ARPES along $\Gamma_2 X_2$ ($k_z$= 0.204~\a) due to their low spectral weight at this $k_z $. However,  these three bands  are clearly visible in the ARPES intensity plot along $\Gamma X$ ($k_z$= 0~\a) in Fig.~\ref{bandsxgz}b and disperse across the \ef\, in splendid agreement with  the EBS along the same direction in Fig.~\ref{bandsxgz}c. At larger $E$   other bands  (numbered as \textit{4-12}) in the EBS  along $\Gamma X$   are  in very good agreement with ARPES (Fig.~\ref{bandsxgz}b). On the other hand, the bands around the $T$ and $B$ crossings both along and perpendicular to the KNL have negligible spectral weight around $\Gamma X$ and so are  observed  neither in EBS  nor ARPES in Figs.~\ref{bandsxgz}b,c.  Similarly, in Fig.~\ref{crossing}a,i  parabolic bands around $E$= 0.3 eV and $E$= 0.5-0.6 eV are not visible in the EBS or  ARPES.  The is caused by the variation of the  spectral weight mentioned in subsection \ref{effbandstr}.  On the other hand,  the band structure shows the crossings formed by band folding  in every BZ, see Supplementary Fig.~S10,%\ref{supp-Raw_band_XS_para}, 
~where the crossing region is highlighted by  red ovals.  This shows the importance of performing ARPES over multiple CDW BZs in the direction of \q\, to decipher  the influence of the CDW on the electronic band structure.

Perpendicular to the KNL along  $\Gamma Z$ ($\Lambda$),  the bands are represented by one dimensional irreps (Supplementary Fig.~S15%\ref{supp-band_connectivity} 
~and Supplementary Table~S3).%\ref{supp-compatibility}). 
~Since the little group along this direction has lesser symmetry than C$ _{2v}$, degeneracy is not enforced. This is shown by  the spin-splitting  along $\Gamma Z$ in the zoomed insets \textit{iv,v} of Fig.~\ref{bandsxgz}a.  This is also true for the bands calculated along various other high symmetry directions (Supplementary Fig.~S17),%\ref{supp-standard_kpath}), 
~where the insets show degeneracy along $\Gamma X$ and splitting along other directions.  In the $\Gamma Z$ direction  no bands  are found to cross  the \ef, and this is corroborated by ARPES as well as EBS. Here,  a hybridization related gap is observed with bands~\textit{13,14} being the highest occupied ones (red double arrow in Figs.~\ref{bandsxgz}b,c).  This gap has been referred to in previous literature as the CDW gap in \lt~\cite{Brouet2008} and other RTe$_3$ systems~\cite{brouet2004, Gweon1998,Komoda2004,Lee2016,	Chikina2022}. A discussion about the variation of the CDW gap of \lt\, with $k_x$ and $k_y$ is provided in the Supplementary Note 4.

\section{Discussions}

In this work, we  show the existence of the recently predicted~\cite{Xie2021} KNL in the CDW state of \lt, along the $\Sigma$ direction with $C_{2v}$ point group  (Fig.~\ref{crossing}o) based on the results of ARPES,   \textit{ab-initio} DFT, EBS obtained by band unfolding and symmetry arguments.   The origin of the KNL is associated with  the time reversal symmetry (\lt\, being non-magnetic in contrast to the other RTe$_3$ materials) as well as the symmetries of the noncentrosymmetric and nonsymmorphic lattice ($C_2$ rotation along $k_x$ and mirror in the $k_x$-$k_z$ plane).  

The crossings of the CDW-induced shadow and the main %(present also in the non-CDW state)  
~bands  enforced by the KNL occur from  $E$\AC0.6 eV, and  disperse in $E$ to traverse  the \ef\, as $k_x$ increases (Fig.~\ref{crossing}n).  The dispersion in $E$ is related to the  dispersion of the main band (the  crossing of the main band with \ef\, moves to larger $k_z$ as  $k_x$ decreases, see Figs.~\ref{effcross}a-d, h-k) coupled with the constraint that  the shadow band is always separated from it by \q.  Other spin degenerate bands also cross the \ef\,  along the KNL. We characterize \lt\, as a  KNL metal in the CDW state based on multiple bands crossing the \ef\,  both along and perpendicular to the KNL.

We have furthermore identified spinless nodal lines   in a general direction on the $k_x$-$k_z$  plane in the limit of negligible SOC (Fig.~\ref{crossing}h). These nondegenerate bands  with distinct irreps involved in the crossings are  linear and traverse the \ef\, with high Fermi velocity. The nodal lines are entirely gapped with SOC.   These characteristics resemble those of  noncentrosymmetric topological nodal line semi-metals  such as  pinictides e.g., CaAgAs~\cite{Yamakage2016},  in which the nodal line is  entirely gapped by SOC resulting in a topological insulator~\cite{Yamakage2016,Lv2021}. Whether the SOC-induced gap in \lt\,  has a topological character is an open question that would require further investigation.  

The appearance of these two distinct types of nodal lines can be attributed to the bilayer splitting, that splits both the shadow and the main bands to bands belonging to different irreps. Thus, the foundation of our investigation is the  unambiguous observation of the bilayer-split  shadow bands by ARPES with which the theory is in impressive agreement.  This occurred due to  the 7 fold modulated structure with \q= $\frac{2}{7}$$c^*$= 0.286$c^*$, employing which DFT was done for the first time   with   \q\,  and the amplitude of the CDW modulation  close to the experimental value. However, the experimental values differ depending on the  method:  for example,  electron diffraction showed \q= 0.28$\pm$0.01$c^*$~\cite{DiMasi1995}, whereas XRD reported \q= 0.2757(4)$c^*$~\cite{Malliakas2006} in the temperature range of 90-100~K. Here, from ARPES we find \q\, to be 0.28$\pm$0.005$c^*$.   Considering these values, it is possible that there might be a minor deviation of  experimental \q\,   from $\frac{2}{7}$$c^*$; hence, is often considered as ``incommensurate"~\cite{DiMasi1995,Malliakas2006}. In order to answer the question whether the KNL would exist if \q\, deviates, we find that it is possible to construct  a 29-fold  (1$\times$1$\times$29) structure with \q= $\frac{8}{29}$$c^*$= 0.2759$c^*$  from the experimental XRD structure~\cite{Capillas2011}.  Note that it has the same space group  (SG\,\#40) as the 7-fold structure.  Its \q\, equals the experimental XRD value, which is most precise among the distinct methods, rounded to the third decimal place. Moreover, the required displacement of the 232 atoms in the primitive unit cell from the atomic positions from  XRD~\cite{Malliakas2006}  is zero  (0.0000) for virtually all the atoms, with the exception of  only a couple of  Te atoms where it is 0.0001\,\AA.  Since  the symmetries of \textit{SG}\,\#40 as well as the TRS are retained in the 29-fold structure, it is expected that the  KNL will remain intact.

In conclusion, the discovery of CDW-induced %nodal lines including
~ KNL in \lt\, -- a non-centrosymmetric, quasi-2D, TRS-preserving material --  is the cornerstone  of our work, which we feel will stimulate further investigation into the fascinating realm of CDW materials.

\section{\small{Methods}}

{\small \noindent\underline{Experimental:} 
 Single crystals of LaTe$_{3}$ with residual resistivity ratio (RRR, $\rho$(300 K)/$\rho$(2 K)) of \AC270 were grown by tellurium flux technique, as discussed in our previous work~\cite{Pariari2021}.

The ARPES measurements presented here were performed at the SGM3 beamline at the ASTRID2 synchrotron facility~\cite{Hoffmann2004}. ARPES data at SGM3 beamline were collected with an energy resolution of 18 meV  at h$\nu$= 24.4 eV.  The  angular resolution was 0.2$^\circ$ (0.008\,\a).  The measurements were performed at 100~K with different photon energies ranging from 13 eV to 140 eV.  All the measurements were done on freshly cleaved surfaces  at a chamber base pressure better than 2$\times$10$^{-10}$ mbar.  The ARPES intensity plots measured as a function of photon energy (h$\nu$) have been converted to $k_{y}$ assuming the free electron final states~\cite{Ngankeu2017}. The raw data ($E_{kin}$ vs h$\nu$) have been converted to $E$ vs $k_{y}$ by utilizing the expression  $k_{y}$= $\frac{1}{\hbar}$$\sqrt{2m(E_{kin}cos^{2}\theta+V_{0})}$,  where $E_{kin}$ is the kinetic energy of the photoelectrons, $\theta$ is the emission angle and the inner potential $V_0$ has been assumed to be 7.5 eV based on the best matching with the ARPES data. The curvature plots have been obtained  as in our earlier work~\cite{Singha2022,Sadhukhan2019} with respect to both \textit{E} and \textit{k} axes  following the method proposed by Zhang $et. al.$\cite{Zhang2011}. This method improves the visibility of the weaker bands in the  ARPES intensity plots and is better than the second derivative approach.  The data analysis has been performed using IGOR pro (version 9).  The dashed curves in Figs.~\ref{effcross}e,f are obtained by a curve fitting the maxima of the MDCs. The maxima are identified by fitting  the MDCs with Lorentzian functions.  The least-square error method has been used for the curve fitting.

The STM measurements were carried out at a base pressure of 2$\times$10$^{-11}$ mbar using a variable temperature STM from Omicron Nanotechnology GmbH, while LEED and preliminary ARPES measurements were carried out in a workstation from Prevac sp. z o.o. STM was performed in the constant current mode using a tungsten tip that was cleaned by sputtering and voltage pulse method. The tip was biased and sample was kept at the ground potential. LEED is performed using a four grid rear view optics from OCI Vacuum Microengineering. Both STM and LEED were performed in the CDW state  at room temperature. 

\vskip 1 \baselineskip
\noindent\underline{Density functional theory:} 
The DFT calculations have been performed for the 7-fold  modulated    structure  with \textit{C2cm} space group (\textit{SG}\,\#40). This structure has been derived from the experimental atomic positions at 100~K  reported in Ref.~\onlinecite{Malliakas2006} using the PSEUDO program with average  displacement of the atoms being $\leq$0.01\AA~\cite{Capillas2011}. VESTA software has been used for Crystal structure visualization~\cite{Momma2011}.

We have employed the DFT-based Vienna Ab-initio Simulation Package(VASP)~\cite{Kresse_1996,Kresse_1999} within the framework of the projector augmented wave method(PAW)~\cite{Kresse_1996,Kresse_1999} to investigate the electronic structure of  LaTe$_3$. The exchange-correlation functional is treated under the generalized gradient approximation(GGA) given by Perdew, Burke, and Ernzerhof~\cite{Perdew}. We have considered 11 valence electrons of the La atom ($5s^25p^65d^16s^2$) and 6 valence electrons of Te atom ($5s^25p^4$) in the PAW pseudopotential. The energy cut-off is set to 500 eV for the expansion of the planewaves. The convergence criterion for energy in the self-consistent-field cycle and total force tolerance on each atom are taken to be 10$^{-6}$ eV and 0.02 eV/\AA, respectively.  The SOC is employed by a second-variation method as implemented in the VASP code~\cite{Kresse_1999}.    To calculate the EBS, we have unfolded the band structure of the CDW state into the primitive BZ of the non-CDW state, using the PyProcar python code~\cite{Herath2020}. All the DFT bands (and  consequently the EBS)  are rigidly shifted to larger $E$ by 0.1 eV  with respect to the \ef\, for comparison with the ARPES data. 

Furthermore, to reconfirm the band crossings and to identify their irreps,  DFT calculations have been performed using the all electron WIEN2k programme package~\cite{wien2k}  and Quantum Espresso  software package~\cite{Baroni_2001}.  Supplementary Fig.~S18%\ref{supp-Band_matching} 
~shows that the bands  are in agreement between the different methods. 
WIEN2k was  performed with energy cut-off of 16 Ry, where the R$_{MT}$K$_{max}$ value is taken to be 9.5. Further, we have used 10 for the maximum value of angular momentum for the (l,m) expansion of wave function or density. Convergence criteria for energy and charge have been taken to be 10$^{-5}$ Ry and 0.001 e$^{-}$, respectively. Quantum Espresso calculations have been carried out using fully relativistic PAW pseudopotentials for La and Te atoms. A planewave cutoff of 80 Ry and a $6\times6\times1$ k-grid were taken along with the energy accuracy of 10$^{-8}$ Ry.

 \section{Acknowledgment}
 S.S., P.S. and S.R.B.  gratefully acknowledge the  financial support from  Department of Science and Technology, Government of India within  the framework of the DST-Synchrotron-Neutron Project to perform experiments at ASTRID2 synchrotron facility.    A part of this work was supported by VILLUM FONDEN via the Centre of Excellence for Dirac Materials (Grant No. 11744). The Computer division of Raja Ramanna Centre for Advanced Technology  is thanked for installing the DFT codes and providing support throughout.

\section{Author contributions}
\small {S.S., P.S., D.C., and S.R.B. conducted the ARPES  measurements with assistance and support from M.B. and P.H.  LEED was carried out by S.S. and P.S., while  STM was performed by V.K.S.  J.B. and R.D. did the DFT calculations under the supervision of A.C.  The  explanation of the results was provided by S.S., T.D.,  A.C., and S.R.B.   The single crystals of \lt\, were grown by A.P., S.R. and P.M, the latter  introduced us to this system.  S.S. analyzed the experimental data with initial help from D.C.,  performed the post-analysis of the DFT results with some inputs from J.B., and  prepared the figures. The project  was planned and led by S.R.B. who wrote the paper  with significant contributions from  S.S., P.H.,  and A.C.   }\\

\section{Competing interests}
The authors declare no competing interests.}

%\clearpage
\newcommand{\noop}[1]{}
\end{document}